\documentclass[11pt,twoside]{article}

\usepackage{asp2006}
\usepackage{epsf}
\usepackage{lscape}

\begin{document}
\title{The Stellar Populations of Lyman Break Galaxies at $z\sim5$}


\author{Kiyoto\ Yabe\altaffilmark{1}, Kouji\ Ohta\altaffilmark{1}, Ikuru\ Iwata\altaffilmark{2}, Marcin\ Sawicki\altaffilmark{3},\\ Naoyuki\ Tamura\altaffilmark{4}, Masayuki\ Akiyama\altaffilmark{4}, and Kentaro\ Aoki\altaffilmark{4}}

\altaffiltext{1}{Department of Astronomy, Kyoto University, Sakyo-ku, Kyoto, 606-8502, Japan}
\altaffiltext{2}{Okayama Astrophysics Observatory, NAOJ, Okayama, 719-0232, Japan}
\altaffiltext{3}{Department of Astronomy and Physics, St. Mary's University, Halifax, B3H 3C3, Canada}
\altaffiltext{4}{Subaru Telescope, NAOJ, 650 North A'ohoku Place, Hilo, HI 96720, USA}

\begin{abstract} 
We present the results of Spectral Energy Distribution (SED) fitting analysis for Lyman Break Galaxies (LBGs) at $z\sim5$ in the GOODS-N and its flanking fields. From the SED fitting for $\sim100$ objects, we found that the stellar masses range from $10^{8}$ to $10^{11}M_{\odot}$ with a median value of $4\times10^{9}M_{\odot}$. By using the large sample of galaxies at $z\sim5$, we construct the stellar mass function (SMF) with incompleteness corrections. By integrating down to $10^{8}M_{\odot}$, the cosmic stellar mass density at $z\sim5$ is calculated to be $7\times10^{6}M_{\odot}\textrm{Mpc}^{-3}$.
\end{abstract}


\section{Introduction}
Recent observations show the gradual increase of the stellar mass density of the universe with time \citep[e.g.,][]{wilkins08}. However, the studies on the stellar mass of galaxies at $z\ga5$ are restricted because the lack of sufficiently deep mid-infrared data. With the advent of Spitzer, we can access the rest-frame optical wavelength, and thus, the stellar mass of galaxies at $z\sim5$. In this work, we explore the stellar mass of Lyman Break Galaxies (LBGs) at $z\sim5$ using the Subaru/S-Cam imaging data and the Spitzer/IRAC data.

\section{Data and Results}
We use the LBGs selected by \citet{iwata07}. Among them, we selected objects which appear to be isolated and not contaminated by neighboring objects in IRAC images with eye inspection. With the archival IRAC images for the GOODS-N and IRAC images we observed in the flanking fields, we construct observed SEDs for the large sample.
We use the SED fitting algorithms based on \citet{sawicki98}. The observed SEDs are compared to the Bruzual \& Charlot 2003 stellar population synthesis models with $0.2Z_{\odot}$ and the Salpeter IMF assuming the continuous star formation history. We use the Calzetti's extinction law. We also consider an effect of H$\alpha$ emission line. The median values of the resulting stellar masses, ages, color excesses, and star formation rates are $4\times10^{9}M_{\odot}$, 25 Myr, 0.23 mag, and $190M_{\odot}\textrm{yr}^{-1}$, respectively. The detailed results are presented by Ohta et al. (2008 this volume).

\begin{figure}
\plotone{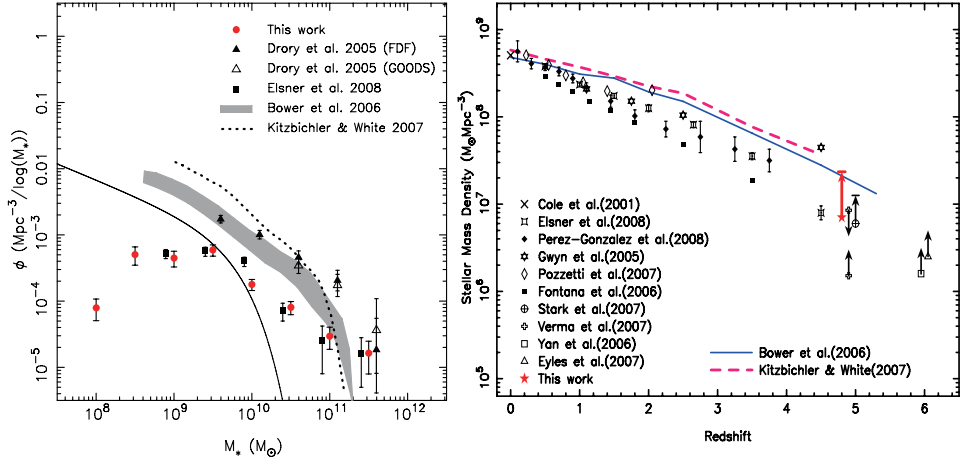}
\caption{Left: Comparison of the stellar mass function of our sample with other observations and theoretical models. Right: The cosmic stellar mass density as a function of redshift. Our result is indicated by a star and the possible upper limit is shown with a horizontal bar. \label{figure}}
\end{figure}

The large sample of LBGs allows us to derive the stellar mass function (SMF). We construct a fiducial SMF after applying incompleteness corrections. The SMF of LBGs at $z\sim5$ is shown in the left panel of Figure \ref{figure}. The SMF of our sample is compared with other observations and theoretical predictions. Although our result disagrees with the result by \citet{drory05}, who only use up to $K_{s}$-band data in the SED fitting, our resulting SMF broadly agree with \citet{elsner08}, who incorporate IRAC data. \citet{elsner08} reported that stellar masses tend to be overestimated without including IRAC data in the SED fitting, especially for $z\ga5$ galaxies. The number densities of our SMF are generally smaller than those of the predictions of semi-analytic models. Our result shows a downsizing phenomenon from $z\sim5$ to $z\sim3$: The number densities in massive end are comparable to some $z\sim3$ results but the number densities in less massive part are smaller than the $z\sim3$ results. By integrating the fiducial SMF down to $10^{8}M_{\odot}$, the cosmic stellar mass density at $z\sim5$ is calculated to be $7\times10^{6}M_{\odot}\textrm{Mpc}^{-3}$. The stellar mass estimation depends largely on the assumed star formation history. Taking into account the uncertainties, we estimate a possible upper limit to be $2\times10^{7}M_{\odot}\textrm{Mpc}^{-3}$. The right panel of Figure \ref{figure} shows that our result broadly agrees with the overall trend of the increase of the stellar mass density with time.  The possible upper limit of our result is almost comparable to predictions by semi-analytic models. However, the models globally tend to overproduce the stellar mass at most redshift.

\end{document}